\documentclass[runningheads]{llncs}
\usepackage{caption}
\usepackage{subcaption}
\usepackage{graphicx}
\usepackage{xcolor}
\usepackage{hyperref}

\usepackage{algpseudocode}
\usepackage{amsmath}

\usepackage{algorithm}

\usepackage{array}
\usepackage{booktabs,makecell,tabularx}

\newcolumntype{C}[1]{>{\centering\arraybackslash}p{#1}}
\newcolumntype{L}{>{\raggedright\arraybackslash}X}

\newcommand{\comment}[1]{}

\usepackage[utf8]{inputenc}

\usepackage{amssymb}

\newcommand{\bx}{\mathbf{x}}

\newcommand{\by}{\mathbf{y}}

\begin{document}
\title{Honeypot Allocation for Cyber Deception in Dynamic Tactical Networks: A Game Theoretic Approach\thanks{Distribution Statement A: Approved for public release. Distribution is unlimited. Research was sponsored by the DEVCOM Army Research Laboratory and was accomplished under Cooperative Agreement Numbers W911NF-23-2-0012 and W911NF-13-2-0045 (ARL Cyber Security CRA). The views and conclusions contained in this document are those of the authors and should not be interpreted as representing the official policies, either expressed or implied, of the Army Research Laboratory or the U.S. Government. The U.S. Government is authorized to reproduce and distribute reprints for Government purposes notwithstanding any copyright notation herein.}}
%
%
\author{Md Abu Sayed\inst{1}\orcidID{0000-0002-5560-9150} \and Ahmed H. Anwar \inst{2}\orcidID{0000-0001-8907-3043}   \and Christopher Kiekintveld\inst{1}\orcidID{0000-0003-0615-9584} \and 
Charles Kamhoua\inst{2}\orcidID{0000-0003-2169-5975}}
\authorrunning{M. A. Sayed et al.}
%
\institute{University of Texas at El Paso, TX 79968, USA \\
\email{msayed@miners.utep.edu, cdkiekintveld@utep.edu} \and
DEVCOM Army Research Laboratory, MD 20783, USA 
\email{a.h.anwar@knights.ucf.edu,charles.a.kamhoua.civ@mail.mil}}
\maketitle              
\begin{abstract}
Honeypots play a crucial role in implementing various cyber deception techniques as they possess the capability to divert attackers away from valuable assets. Careful strategic placement of honeypots in networks should consider not only network aspects but also attackers' preferences. The allocation of honeypots in tactical networks under network mobility is of great interest. To achieve this objective, we present a game-theoretic approach that generates optimal honeypot allocation strategies within an attack/defense scenario. Our proposed approach takes into consideration the changes in network connectivity. In particular, we introduce a two-player dynamic game model that explicitly incorporates the future state evolution resulting from changes in network connectivity. The defender's objective is twofold: to maximize the likelihood of the attacker hitting a honeypot and to minimize the cost associated with deception and reconfiguration due to changes in network topology. We present an iterative algorithm to find Nash equilibrium strategies and analyze the scalability of the algorithm. Finally, we validate our approach and present numerical results based on simulations, demonstrating that our game model successfully enhances network security. Additionally, we have proposed additional enhancements to improve the scalability of the proposed approach.

\end{abstract}
%
%
%
\section{Introduction}\label{sec:intro}
The cybersecurity domain encounters numerous complex issues due to the dynamic nature of threats and the intricate decision-making processes involved. One of the most powerful threats in cybersecurity is the Advanced Persistent Threat (APT) attack where attackers carry out highly targeted, long-term, stealthy attacks against government, military, and corporate organizations \cite{center2013apt1}. In many instances, APTs manage to establish a persistent and concealed presence within a targeted network for extended periods, sometimes lasting months or even years, without being detected.

Additional challenges include protecting dynamic and diverse mobile networks from intense, short attacks such as denial of service(DoS), especially in the context of the Internet of Battlefield Things (IoBT) \cite{abuzainab2017dynamic} and tactical dynamic networks \cite{burbank2006key}. The attackers rely on lateral movement to utilize the network resources in reaching their targets. To counter this, defenders can employ appropriate actions to effectively detect and mitigate lateral movement. In this context, we consider cyber deception via honeypots to proactively mislead attackers.


Computer networks face several challenges that can impact their performance, security, and reliability. The rapid expansion of wireless networking has introduced numerous challenges, including network scalability, resource allocation, interference mitigation, and security. Software-Defined Networking (SDN) encounters new challenges and opportunities in networks, specifically regarding the functionality, performance, and scalability of SDN in cloud computing, IT organizations, and networking enterprises \cite{jammal2014software}.

\comment{bertino2021computing,}
Mobility is a significant characteristic of tactical networks, which introduces distinct challenges such as intermittent network connectivity, temporary power loss, and communication issues. These challenges can be particularly problematic when multiple autonomous computers communicate through a network and interact with each other \cite{nsfsecurity}. As a result, fixed deception policies are suboptimal since one needs to consider the connectivity of the computer network. In our work, we primarily focus on modeling cyber deception in dynamic tactical networks.



Cyber deception represents an advanced proactive technology in the field of cyber defense. Its purpose is to provide attackers with credible yet misleading information, effectively leading them astray. While deception techniques have traditionally been employed in the physical domain as a tactic of traditional warfare, their application has extended to the realm of cyberspace, serving as a means of intrusion detection and defense. In many ways, cyber deception shares similarities with non-cyber deception, encompassing comparable philosophical and psychological characteristics. Proactive measures can be employed with the objective of capturing the attackers and closely monitoring their actions. Honeypots play a vital role in this process, acting as simulated entities within the system or network to deceive the attacker. By studying the attacker's strategies and intentions through the use of honeypots, defenders can enhance their comprehension of the attack and subsequently develop more effective deception schemes \cite{wang2018cyber}.

Honeypots play a crucial role in the realm of cyber deception by serving as effective tools to mislead and divert attackers while consuming their valuable resources. These deceptive elements can be categorized into two types: low-interaction honeypots and high-interaction honeypots. Low-interaction honeypots simulate specific services and are typically implemented in a virtualized environment, offering a relatively simpler setup and operational process compared to high-interaction honeypots. However, it is important to note that low-interaction honeypots are more prone to detection by adversaries, making them easier to identify and bypass \cite{mokube2007honeypots}.



A key challenge in securing tactical networks lies in their mobile nature, rendering the defender's base policy ineffective and sub-optimal over time. For instance, defender honeypot allocation based on the initial network is not useful as network connectivity changes over time as well as not optimal over whole state space. Despite the increasing attention given to cyber deception in the past decade, there remains a gap in the literature regarding its incorporation of mobility features and anticipation of the future evolution of tactical networks. In this paper, we utilize dynamic attack graphs and game theory to model mobility in tactical networks. Specifically, when network mobility is present, the connectivity of the corresponding attack graph undergoes changes in specific edges, thereby redefining potential attack paths and possibly rendering some defender strategies ineffective. To the best of our knowledge, this framework represents a novel approach for proactive defense in the presence of network mobility. 


This paper presents a novel approach for dynamic cyber deception via strategic honeypot allocation given a limited deception budget. By leveraging a game theoretic framework, our objective is to devise an effective honeypot allocation policy throughout the network attack graph that takes into account future changes in network connectivity. We model this problem as a two-player Markov game. In our analysis, the defender anticipates future network mobility. We assume a well-known attacker performed reconnaissance before launching this attack. The proposed model takes into account different node values that reflect the significance of each node in the network.  

We design the state space according to potential changes in network connectivity assuming that a mobile node may lose communications with its neighbors inducing a transition to a new state. As shown in the results sections, it is beneficial for the defender to allocate honeypots according to the current network topology while taking into account the potential transitions to new topologies as well to reduce the cost of reconfiguration in the future. This results in a Markov game model that can be solved via standard Q-minimax algorithm \cite{littman1994markov}. We validate the efficiency of the proposed algorithm that showed a substantial improvement in mitigating attacker impact via deception using strategic allocation for honeypots.
We balance the need for future look-ahead transitions and our numerical results show a faster convergence rate with a reasonable amount of iterations. We compare our defensive deception strategies to other allocation policies. Moreover, we demonstrate that our approach exhibits greater improvement against a less informed attacker that fails to anticipate future transitions. This finding underscores the significance of cyber deception in enhancing network security.

\noindent
We summarize our main contribution below:
\begin{itemize}
\item We design a dynamic game between defender and attacker, to generate a cyber deception strategy against lateral move attacks leveraging attack graphs under network mobility. The game is played on attack graphs that capture vulnerabilities, node importance, and network topology.
\item We design a realistic set of future states considering the possibility of different node losing communications.
\item We model our predictive model in the transition matrix of the Markov game model and solve for the stationary Nash equilibrium strategy at different states. 
\item Finally, we present numerical results that show the effectiveness of cyber deception as well as the fast convergence of the game solver with the presence of network mobility. We evaluated our approach under symmetric and asymmetric information between both players and analyzed the scalability of our approach under different assumptions.
\end{itemize}

The rest of the paper is organized as follows. We describe related work in section \ref{sec:relwork}. In section \ref{sec:model}, we discuss the system model, define the game model, and propose our deception approach. In section, \ref{sec:method} we present the methodology of network mobility-assisted cyber deception. Our numerical results are presented in section \ref{sec:results} and in section \ref{sec:con} we conclude our work and discuss the potential future extension of our research.

\section{Related work}\label{sec:relwork}
Our research builds upon existing work on cyber deception and games on attack graphs to model lateral movement attacks and characterizes game-theoretic deception strategies with the presence of network mobility.

\subsection{Attack Graph:} Attack modeling techniques, such as attack graphs (AGs), provide a graph-based approach to representing and visualizing cyber-attacks on computer networks \cite{lallie2020review}. However, the scalability of generating attack graphs poses a significant challenge, with existing works struggling to handle large enterprise networks  \cite{ou2006scalable}. In this study, we focus on a simplified attack graph where nodes represent vulnerable hosts and edges represent specific exploits for attacker reachability. While this model may not capture every vulnerability, it effectively demonstrates potential attack paths that adversaries can exploit, which is crucial for generating optimal honeypot allocation policies. It is important to note that attack graphs are limited in their ability to directly model mobility, but effectively modifying attack graphs can be used to model mobility in tactical  networks.


\subsection{Game Theoretic Deception:} In cybersecurity research, game theoretic defensive deception has been extensively addressed. Schlenker et al. \cite{schlenker2020game} propose a deception game for defender who decides on deception in response to the attacker's observation while the attacker is either uninformed of the deceit or aware of it. The comprehensive game model of hypergame theory \cite{Fraser84} has been used to simulate the many subjective viewpoints of participants in uncertain situations. \comment{Cho et al. \cite{cho2019modeling} use hypergames to assess the degree to which a defensive deception signal may persuade an attacker}. Wan et al. \cite{wan2021foureye} discuss hypergame-based deception against advanced persistent threat attacks performing multiple attacks performed in the stages of cyber kill chain. Sayed et al.  \cite{sayed2022cyber} propose a game theoretic approach for zero-day vulnerability analysis and deceptive mitigation against zero-day vulnerability. Zhu et al. \cite{zhu2021survey} discuss the synergies between game theory and machine learning \cite{mahmud2023machine,raju2020predicting} to formulate defensive deception. In this paper, we extend cyber deception under network mobility.

\subsection{Network Deception:} 
Computer network deception research focuses on developing techniques, strategies, and technologies to enhance the effectiveness of deception as a proactive cyber defense. Lu et al. \cite{lu2020cyber} describe the  fabrication or manipulation of network-level information such as network topology, host information, tarpits, and traffic information.
Chiang et al. \cite{chiang2018defensive} discuss the use of defensive cyber deception to enhance the security, dependability of network systems, and focus on the application of Software-Defined Networking (SDN). Urias et al. \cite{urias2016gathering} discuss the use of computer network deception as a means to gather threat intelligence. \comment{The paper highlights the changing threat landscape and the need for effective techniques to counter attacks. It explores the concept of using deception to influence the attacker's target selection process.}

\subsection{Mobility in Tactical Network}
In tactical networks, mobility refers to the ability of devices or agents to move within the network while maintaining connectivity and resource access. Mobility involves features like protocol support, seamless roaming, handover management, and location tracking. However, it presents challenges such as intermittent connectivity, location management, handover \& roaming efficiency, quality of service, security \& privacy, and scalability. Overcoming these challenges is essential to achieve uninterrupted connectivity and adaptability in various environments \cite{nsfsecurity}. Mobility also does not follow the same pattern as traffic expansion \cite{sayed2017understanding}.

The characteristics of mobility in tactical networks include various aspects related to the movement and connection of users or nodes within network. Pirozmand et al. \cite{pirozmand2014human} examined human mobility in terms of its geographical, temporal, and connectivity properties. They explore mobility models, traces, and forecasting methods to give a thorough picture of how nodes move within networks. Abdulla et al. \cite{abdulla2007characteristics} analyzed the mobility characteristics of commonly used models, focusing on inter-contact times and the approximation of exponential distributions in opportunistic network scenarios.

The impact of node mobility in tactical networks is a crucial area of research, focusing on how the movement of nodes within a network affects various network characteristics and performance.
Fu et al. \cite{fu2021exploring} investigated the impact of node mobility on cascading failures in spatial networks. This includes studying the influence of node mobility on network load redistribution, the cascading process, and the robustness of network configurations against cascading failures. Xia et al. \cite{xia2012mitigating} proposed a a cluster-based routing protocol called FASTR to mitigate the impact of node mobility in networks with high node mobility and low group mobility. This protocol utilizes a mobile backbone to address the challenges associated with node mobility. 

The mobility model and its parameters significantly impact network communication in wireless mobile opportunistic networks. Various mobility models are proposed to describe random movement patterns of nodes in ad hoc networks, emphasizing the importance of considering node mobility in network design and analysis \cite{lin2015impact}. Pala et al. \cite{pala2015effects} investigate how node mobility influences energy consumption and network lifetime. Results show that mobility can improve energy balancing up to a certain level, but excessive mobility may degrade energy balancing in wireless networks.

Urias et al. \cite{urias2017technologies} highlighted the limited number of deception platforms that have been successfully shown to enable strategic deception in computer network operations environments. This indicates that the development of specific tools and techniques for combining network mobility and cyber deception may still be an area of ongoing exploration. Therefore, network mobility can be incorporated in designing deception techniques such as dynamic movement and placement of deceptive elements within the network.

In our work, we develop deception techniques against lateral movement attacks considering network mobility. This is the first model that explicitly considers the impact of network mobility in designing proactive game-theoretic policies for optimizing deception resources. We present our system model and game formulation in the following sections.

\section{System Model}\label{sec:model}\label{sec:model}

\subsection{Attack Graph Model}\label{subsec:model1}

We consider a targeted attack that follows a thorough reconnaissance phase, during which the adversary gathers all the necessary information about the network structure, node properties, and existing vulnerabilities. To represent these features, we adopt a modified version of the attack graph model \cite{miehling2015optimal} denoted as $G_1(\mathcal{N},\mathcal{E}, \mathcal{\theta}, \mathcal{V})$ where

\begin{itemize}
\item $\mathcal{N}$ represents the set of nodes
\item $\mathcal{E}$ represents the set of edges
\item $\mathcal{\theta}$ represents the set of node types. We assume that each non-leaf\footnote[1]{Non-leaf nodes can be predecessor of at least one node.} node can have two  types such as $\land$ (AND), $\lor$ (OR).
\item $\mathcal{V}$ represents the value associate with each node

\end{itemize}

The set of nodes is interconnected through the set of edges, which depict their accessibility and network connectivity. The defender classifies the nodes based on their importance, functionality, and role through a course of an attack action. Within this classification, there are two distinct subsets of nodes, the set of entry nodes $E$ and the set of target nodes $T$. The remaining nodes are intermediate nodes that an attacker must compromise while progressing from an entry node (attack start node $\in E$) toward a target node $\in T$. The defender decides what nodes are most valuable and are more likely to be targeted and hence labels them as targets either based on previous attack reports or according to expert decisions.

In this graph, each node represents a host that has one or more vulnerabilities that can be exploited to reach a neighboring node. The edge represents the connection that can be utilized by malicious users to reach the next targeted node. A legitimate user at node $v$ possesses the appropriate credentials to access node $u$. However, an adversary can only reach node $u$ by exploiting a vulnerability. At the same time, there must exist an exploitable vulnerability at $u$, an open port at $v$ and node $v$ is reachable via a communication link from node $u$. Such exploitation possibility is represented by $e_{u,v} \in \mathcal{E}$. Each node $i \in \mathcal{N}$ is assigned a value $\mathcal{V}(i)$ which denotes node importance. Therefore, $G_1(\mathcal{N},\mathcal{E}, \mathcal{\theta}, \mathcal{V})$ represents the attack graph, which is assumed to be known to both the defender and the attacker.


The attack graph model also considers node types $\mathcal{\theta}$. The types of nodes, denoted by $\land$ (AND) or $\lor$ (OR), determine the conditions under which a node can be controlled by an adversary. If a node is marked with $\land$ (AND), it means that all of its predecessor nodes must be controlled by the adversary for the node itself to come under adversary control. On the other hand, if a node is marked with $\lor$ (OR), it means that only one of its predecessor nodes needs to be controlled by the adversary for the node to come under adversary control.

In the case of computer network attacks, each adversary operates independently, and exploits a set of nodes to reach a specific target. Therefore, all the nodes within the attack graph are designated with the $\lor$ (OR) node type. 


\begin{figure*}[h!]%
    \centering
    \begin{subfigure}{.40\columnwidth}
        \includegraphics[width=\columnwidth, height=4cm]{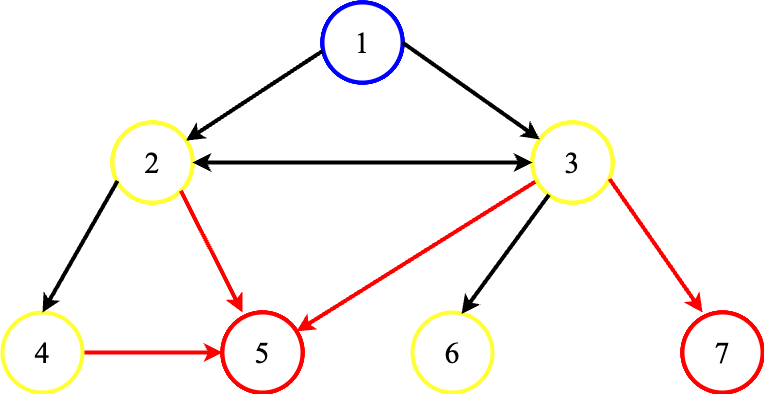}%
        \caption{Initial network with 7 nodes.}%
        \label{fig:qlearn1}%
    \end{subfigure}\hfill%
    \begin{subfigure}{.40\columnwidth}
        \includegraphics[width=\columnwidth, height=4cm]{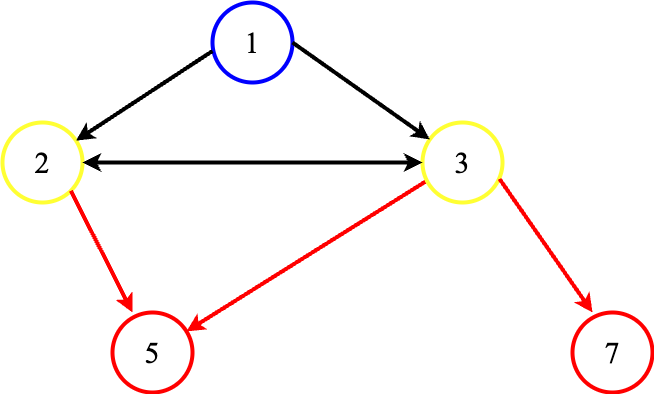}%
        \caption{Network after mobility with 5 nodes.}%
        \label{fig:qlearn2}%
    \end{subfigure}\hfill%
    \caption{7-node tree network topology with a single entry node and two target nodes (5,7) and because of node mobility, node 4 and node 6 will likely abandon the group in the future.}%
    \label{fig:system}%
\end{figure*}

Node mobility denotes node removal on the computer network due to multiple factors including hardware failure, network maintenance, network redesign, security concern, decommissioning, network upgrade, and network optimization. In a tactical network, if all nodes are moving in the same direction and speed which allows them to maintain communications the attack remains static. However, due to specific tactical requirements, one or more nodes may be assigned to change their course and go in other directions resulting in a new attack graph. Removing a node eliminates all exploitable vulnerabilities of that node. In other words, it removes all edges connected to it. We consider a complete information structure where the defender and attacker can observe node mobility in the network. Hence, both players can update the attack graph of the game.   

Fig.~\ref{fig:system} represents a 7-node tree attack graph consisting of one entry node (1), four intermediate nodes (2, 3, 4, 6), and two target nodes (5, 7). In this network, there is one available path for reaching target node (7), while there are three paths for reaching target node (5). Due to node mobility, nodes 4 and 6 are likely to lose communications in the future, leaving only two available attack paths to reach target node (5). Future mobility information should impact the initial honeypot allocation strategy deployed by the defender from the beginning. Our model quantifies and captures the advantages of considering the network's future evolution. In this example, we consider nodes leaving the network, however, the proposed model is general to adapt any future topologies including adding or removing new connections, or new nodes. 

\subsection{Defender model}
At a given state, the defender strategically places a set of honeypots along the network edges among the set of edges leading to the set of target nodes $T$ to deceive the attacker. The honeypot budget is denoted as $H$. Hence, the defender's action space, denoted as $\mathcal{A}_d$, consists of possible allocation vector $\mathbf{e}$ in which $\mathcal{A}_d = \{ \mathbf{e} \in 2^{\mathcal{E}} \mid \mathbf{e}^T1 \leq H \}$. Where, $\mathbf{e}$ is a binary vector of length $|\mathcal{E}|$, where each entry $\mathbf{e}(i)$ equals 1 if a honeypot is allocated along the $i^{th}$ edge, and is set to zero otherwise. 
To balance the defender strategies, the defender pays a cost, $C_d$, per each installed honeypot in the network, otherwise, the defender will always try to maximize the number of allocated honeypots. The total cost can be expressed as $C_d \times \left| a_d\right|_1$, where $ \left| a_d\right|_1$ is the number of honeypots associated with the action $a_d$. Finally, assuming both players are rational, the defender aims to reduce the attacker's reward by placing honeypots on edges that are attractive to the attacker, while minimizing the total cost of the played deception strategy.
\subsection{Attacker model}
The defender considers a practical scenario where the attacker had gathered valuable reconnaissance information about the network topology before launching this attack. Hence, the attacker is launching a targeted attack to compromise a specific subset of nodes, $T$. Therefore, she selects one of the possible attack paths to reach a target node to maximize its expected reward. Thus, the attacker's action space, denoted as $\mathcal{A}_a$, consists of all possible attack paths starting from an entry node $u \in E$ to a target node $v \in T$. The attacker incurs an attack cost that depends on the selected attack path. We consider a cost due to traversing a node in the attack graph denoted by $C_a$. The attacker faces a tradeoff between traversing important nodes while reducing his overall attack cost. 

\subsection{Reward function}
 
The reward function is formulated to capture the tradeoff that faces each player. For each action profile played  $(a_d, a_a) \in \mathcal{A}_d \times \mathcal A_a$, the defender receives a reward $R_d(a_d, a_a)$ and the attacker reward is $R_a $. We consider a zero-sum game where $R_a + R_d =0$.
Recall that each node $i \in \mathcal{N}$ is assigned a value $v(i) \in \mathcal{V}$ that reflects its importance, the defender gains more by protecting high-valued nodes via correct placement of honeypots. On the other hand, the attacker reward increases when attacking nodes of high values along the selected attack path while evading honeypots.
 
The defender reward is expressed as:


\begin{align} \label{eq:1}
      R_d(a_d, a_a) = \sum_{i \in a_a} \left[Cap \cdot v(i)\cdot \mathbf{1}_{\left\{ i \in a_d\right\}} \nonumber  -Esc \cdot v(i)\cdot \mathbf{1}_{\left\{ i \notin a_d\right\}}\right]\nonumber & \\ - C_d \cdot \left\| a_d\right\|_1 + C_a (a_a)
\end{align}

Here, $Cap$ represents the capture reward received by the defender when the attacker encounters a honeypot along the selected attack path $a_a$. On the other hand, $Esc$ denotes the gain for the attacker upon a successful attack from one node to another while progressing toward the target node.
 
Finally, $C_d$ and $C_a(a_a)$ are the cost per honeypot, and attack cost, respectively. The attack cost is proportional to the length of the attack path as the attacker could become less stealthy due to numerous moves.

Now we define a two-player zero-sum game for a particular state, $s$, $\Gamma(s) (\mathcal{P}, \mathcal{A}, \mathcal{R})$, where $\mathcal{P}$ is the set of the two players (i.e., defender and attacker). The game action space $\mathcal{A} = \mathcal{A}_d \times \mathcal{A}_a$ as defined above, and the reward function $\mathcal{R}=(R_d, R_a)$.

The finite game developed above admits at least one NE in mixed strategies \cite{basar}.
Let $\mathbf{x} $ and $\mathbf{y} $ denote the mixed strategies of defender, and attacker when the game is played on graph, $G$. 
The defender expected reward of the game can be expressed as:
\begin{equation}
    U_d(G)=\mathbf{x}^TR_d(G)\mathbf{y}
\end{equation}
where $R_d(G)$ is the matrix of the game played on $G$ and the attacker expected reward $U_a(G) = - U_d(G)$. Both defender and attacker can obtain their NE mixed strategies $\mathbf{x}^*$ and $\mathbf{y}^*$ via a linear program (LP) as follows,
\begin{equation}
\begin{aligned}
&\underset{\mathbf{x}}{\text{maximize}}
& U_d \\
& \text{subject to}
& \sum_{a_d \in \mathcal{A}_d} R_d(a_d,a_a){x}_{a_d} \geq U_d, & \;\; ~~~~ \forall a_a \in \mathcal{A}_a.\\
& & \sum_{a_d \in \mathcal{A}_d} {x}_{a_d} =1,~~ & {x}_{a_d} \geq 0,
\end{aligned}
\label{eq:global_LP}
\end{equation}

\noindent
where ${x}_{a_d}$ is the probability of taking action $a_d \in \mathcal{A}_d$.  

Similarly, the attacker's mixed strategy can be obtained through a minimizer LP under $\mathbf{y}$ of $U_d$. 

\section{Dynamic Game Model}\label{sec:method}
In the previous section, we show the formulation of one stage game. In this section, we extend the formulation for a dynamic muti-stage game (Markov game) between the defender and the attacker due to network mobility. 

In a dynamic environment, the game is played under varying circumstances each time, encompassing different network connectivity configurations, changes in connectivity, and as well as patching existing vulnerabilities. In our work, we primarily focus on the mobility of the network over time. To comprehensively analyze the progression and evolution of this game, we employ a Markov game framework where the state of the game captures all information needed to generate a honeypot allocation strategy.
We assume that the defender changes the allocations based on the new topology. Players reward is the total reward over all future states.

Let $s$ denote the state of the game defined as the attack graph associated with the network topology at state $s$. A dynamic game $\Gamma$ is defined as the tuple $(\mathcal{K,A,S,P,R})$, where $\mathcal{K}$ is a set of two players, $\mathcal{S}$ is a finite set of states. We consider an uncontrolled dynamic game where  $\mathcal{P}: \mathcal{S \times S'} \to [0,1] $ is a transition probability function between states such as $P(s, s')$ denotes the probability of transitioning from state $s$ to the future state $s'$. The action space $\mathcal{A} = \Pi_{s \in \mathcal S} \mathcal{A}(s) $ and reward function $\mathcal{R}$. Each player aims to maximize his long-term expected payoff. Where $R(s')$ is the immediate reward as defined in Section \ref{sec:model} for any state $s'$. A terminal state $s'' \in \mathcal{S}$ is a state, where no transition future transitions can be reached from $s''$. In other words, at any terminal state, players receive immediate rewards onward.

\subsection{State Space and Game Transitions}
Given an initial attack graph (full topology),  network mobility may induce new connections, and/or result in removing nodes from the network topology. As we discussed in the Section \ref{sec:intro}, node mobility in tactical networks renders the defender's base strategy ineffective, and requires reconfigurations to the initial honeypot allocation strategies as well as other security resources in the network. 

For practical constraints, it is difficult for the defender to anticipate new connections to be added to the initial attack graph due to mobility from the initial topology. Therefore, and without loss of generality, our model does not consider transitioning to such states. Additionally, we focus on removing connections due to losing communications between mobile nodes/agents. The defender builds this model using information regarding node movement patterns in the future. For simplicity, we consider one node change at a time. In other words, $\mathcal{N}(s) - \mathcal{N}(s') = u$, where $u$ is a node moving away from the tactical network set of nodes/agents. To formally define our Markov game, the defender needs to compute the transition probability matrix, $\mathcal{P}$. When node $u$ moves away, it is removed from the attack graph state, and hence we transition to a new state $s'$. The transition probability, denoted as $P(s,  s')$, is defined as $1 - (\text{value proportion} \times \text{degree proportion})$ \begin{equation}\label{eqn:probability}
 p(s,s') = 1 - \frac{\nu(u)}{\nu_{\max} } \times \frac{\delta(u)}{\delta_{\max}},   
\end{equation}
\noindent
where $ 0 < \nu(u) < \nu_{\max} $ is the value assigned to node $u$, $\delta(u)$ is the degree of node $u$ where $\delta_{\max}$ is the max node degree in the network. The formulation in (\ref{eqn:probability}) follows a practical assumption that high-valued nodes and central nodes in the network are less likely to abandon the tactical network. However, a node with less connectivity and leaf nodes will have a higher probability of being disconnected. Additionally, we assign a non-zero
probability for no state change (i.e., self-transition). Hence, $p(s,s) = \mu $, and $ \sum_{s' \neq s} p(s,s') = 1- \mu ~, ~ \forall s \in \mathcal{S}$, such that $\mu$ is the chance of experiencing no mobility. This fully defines the dynamic game model. 

\subsection{Nash Equilibrium Analysis}

Let $x^i(s);~ i = 1,\ldots, |\mathcal{A}_d|(s)$, and $y^j(s);~ j = 1,\ldots, |\mathcal{A}_a|(s)$, denote the probability that the defender and attacker play the $i^{th}$ and $j^{th}$ pure actions at state $s$ from their corresponding available actions spaces, $\mathcal{A}_d (s)$ and $\mathcal{A}_a (s)$, respectively.   
A stochastic stationary policy is readily defined over all states as $\pi_d = \left \{ \bx(s^1),\ldots,  \bx(s^n)\right \}$ for the defender, and $\pi_a = \left \{ \by(s^1),\ldots,\by(s^n)\right \}$ for the attacker, where $n=|\mathcal{S}|$ is the total number of states. 






Each player can maximize their expected reward by greedily maximizing the $Q(s)$-function at each state which is defined below in equation (\ref{eqn:Q-value}). 
The main goal of the defender is to maximize the expected discounted rewards. Under some stationary defense and attack policies, $\pi_d$ and $\pi_a$ respectively, the expected sum of discounted rewards starting from some initial state $s \in \mathcal{S}$ at time $t = 0$ is given by:
\begin{align}\label{eqn:V}
V(s,\pi_d,\pi_a) = \mathbb{E}\left[\sum_{t = 0}^\infty \gamma^t R_d\left(s_t,a_d,a_a,s_{t+1}\right) \Big| s_0 = s,\pi_d,\pi_d\right],
\end{align}
It is worth noting that, the immediate reward $R_d(.)$ depends not only on the current state $s$ but on the future state $s'$ as well. The expectation in (\ref{eqn:V}) is taken over the players' stationary policies (noting that $\pi_d$ and $\pi_a$ are randomized policies) and the state evolution, denoted as $\mathcal{P}(.)$. The subscript $t$ represents the $t$-th stage, and $0 < \gamma < 1$ is a discount factor. Based on the findings in Section \ref{sec:model}, this finite game achieves a value $V(s)$ at each state $s$. Moreover, there exists a mixed strategy Nash equilibrium (NE) where $V^(s)$ denotes the value under the equilibrium mixed strategies $\pi^*_d(s)$ and $\pi^*_a(s)$ at this state. To determine the optimal stationary policies for both players (i.e., NE), can be learned using value iteration over the value function at each state which is equivalent to finding the value of the game which is defined as:

\begin{align}\label{eq:optimal_policy}
V^*(s) := V(s,\bx^*(s)^*,\by^*(s)^*) = \max_{\bx(s)}\min_{\by(s)}  V(s,\bx(s),\by(s))~~~;~~~ \forall s \in \mathcal{S}
\end{align}




\noindent
The optimal randomized stationary policies $\bx^*(s)$ and $\by^*(s)$ for state $s$ are the solutions to the following equation:
\begin{equation}\label{eqn:V_opt}
 V^*(s) =    \max_{\bx(s)}\min_{\by(s)}   ~\mathbb{E}\left[R(s,a^{(d)},a^{(a)},s') + \gamma V^*(s') \Big| \bx(s),\by(s)\right],
\end{equation}
\noindent
where the immediate reward term is given by
\begin{align}
& \mathbb{E}\left[R(s,a_d,a_a,s')\Big| \bx(s),\by(s)\right] =    \sum_{s'\in\mathcal{S}}\sum_{a_d \in \mathcal{A}_d}\sum_{a_a \in \mathcal{A}_a}R(s,a_d,a_a,s') x(s)y(s) p(s',s)
\end{align}
Thus, the optimal stationary policies are 
$\pi_d^* = \{\bx^*(\mathsf{s}^1),\ldots, \bx^*(\mathsf{s}^n)\}$, $~\pi_a^* =\{\by^*(\mathsf{s}^1),\ldots, \by^*(\mathsf{s}^n)\} $.

A Dynamic game can be solved using value iteration following Markov Decision Process (MDP) intuition \cite{littman1994markov}. In this context, the value of state $s$  by solving:
\begin{equation} \label{eq:maxQ}
V(s)= \max_{\pi_d \in \Delta(\mathcal A_d)} \min_{a_a \in \mathcal A_a} Q(s,a_d,a_a),
\end{equation}
where $Q(.,a_d,a_a)$ denotes the quality of the state-action pair defined as the total expected discounted reward achieved by following a non-stationary policy that takes action $a_d$ and then continues with the optimal policy thereafter. For a dynamic game, the state-action quality function is defined as:
 
\begin{equation}\label{eqn:Q-value}
Q(s,a_d,a_a) =   \mathbb{E} \left [ R(s,a_d,a_a,s')  + \gamma V^*(s') \Big| a_{d},a_a\right] 
\end{equation}
\noindent
We solve (\ref{eqn:V_opt}) to find the optimal policy, by iterating over the value function following the Q-minimax algorithm introduced in \cite{littman1994markov}.  
 
\begin{equation} \label{eqn_max}
V^*(s)= \max_{\bx(s) \in \Delta(\mathcal {A}_d)} \min_{a_a \in \mathcal {A}_a} \sum_{a_d \in \mathcal{A}_d} Q(s, a_d, a_a)  x^{a_d}(s),
\end{equation}

The tabular nature of the Q-minimax algorithm evidently faces the curse of dimensionality due to the size of the state-action space. To enhance its scalability, we experiment with various techniques, such as limiting the state space to include future states that impact the attack paths leading to target nodes. In the following section, we present our numerical results and compare the performance of the learned policies derived from our game model with other deceptive policies.

\begin{algorithm}
\caption{Predictive Model (Proposed Algorithm)}
\begin{algorithmic}[1]
\Procedure{Input}{{Topology, H, Esc, Cap, $C_d$, $C_a$, $\mathcal{V}$}}
    \State Initialize: $\mathcal{S}$, s0, entry node, E, T
    \State $h = 1$ \;
    \While{$h < \ell$}
        \State Generate all new nodes at depth $h$
        \State Define and solve all static game at depth $h$
        \State $h++$
    \EndWhile
    \State $h = l - 1$ \;
    \While{$h > 0$}
        \State Calculate $(\bx^*(\mathsf{s}^h), \by^*(\mathsf{s}^h))$
        \State $h--$ \;
    \EndWhile
    \State \textbf{Return} $(\pi_d^*, \pi_a^*)$
\EndProcedure
\end{algorithmic}
\end{algorithm}

\section{Numerical Results}\label{sec:results}

In this section, we analyze and validate our game-theoretic model by examining the obtained numerical results. Specifically, we evaluate the convergence, effectiveness, and scalability of our approach. Firstly, we ensure and demonstrate the convergence of the extended Q-Learning algorithm to optimal (Nash equilibrium) strategies as well as the convergence of both the state value function as well as the rate of convergence at different networks. Secondly, we evaluate the performance of the developed approach by comparing the attacker's reward against various deception policies, including random, myopic (which ignores future transitions), and our policy based on the proposed predictive model. We consider formulation including full state space  and compact state space. Lastly, we assess the scalability of our algorithm by comparing its performance on both full state and compact state space formulations.

\begin{figure}[h!]
    \centering
    \includegraphics[width=12cm, height=5cm]{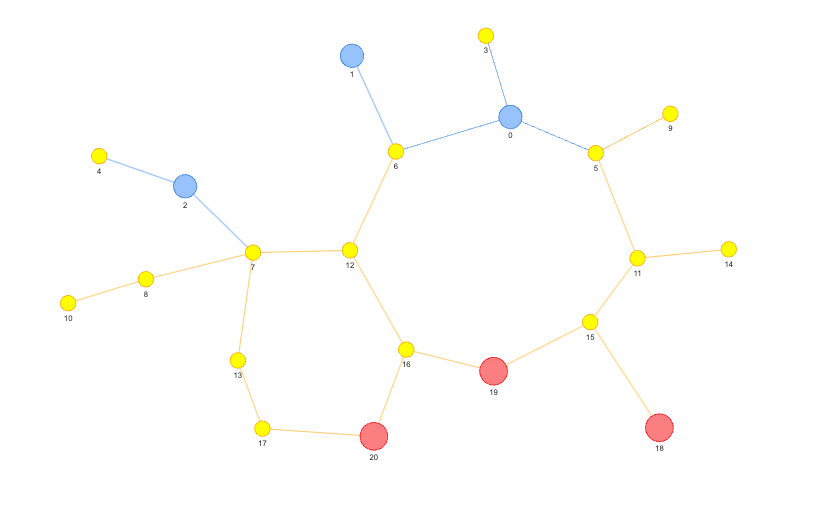}
    \caption{Network topology consists of 20 nodes, with entry nodes represented in blue, target nodes in red, and intermediate nodes in yellow.}
    \label{fig:network}
\end{figure}


We analyze the potential impact of network mobility by generating attack graph by NetworkX library \comment{\cite{hagberg2008exploring}} and the attack graph also follows the definition in section \ref{subsec:model1}. We identify the subsets of entry and target nodes. Each intermediate node is assigned a value generated randomly between 10 and 50. Our 20-node network topology has 3 entry nodes (in blue) and 3 target nodes (in red) as shown in Fig. ~\ref{fig:network}. For this 20-node network, the defender has 10 actions for honeypot allocations including (6,12), (5,11), (16,19), (11,15), (0,5), (16,20), (12,16), (0,6), (15,19), (15,18), while the attacker selects between 4 attack paths: path1 = [0, 5,11,15,18], path2 = [0,5,11,15,19], path3 = [0,6,12,16,19], path4 = [0,6,12,16,20]. 


To incorporate network mobility, we remove nodes from the attack graph, leading to transitioning to a new state and having a modified attack graph. We then explore how to update the defender's base policy by considering both future rewards and transitions. To address the dimensionality of the Q-minimax algorithm, we adopt a formulation with a two-step look-ahead. Additionally, we explore two different state-space representations. The first is a full state space that encompasses all possible transitions, where any node in the intermediate subset can leave the network, resulting in a new state transition. The second is a compact state space that includes only future states associated with nodes belonging to attack paths. The compact state representation is based on the intuition that the mobility of nodes that do not belong to any attack path will have no impact on the optimality of the defender's strategies. For the 20-node network, the full state space consists of 272 states, while the compact state space has 19 states.

\begin{figure*}[h!]%
    \centering
    \begin{subfigure}{.50\columnwidth}
        \includegraphics[width=\columnwidth, height=5cm]{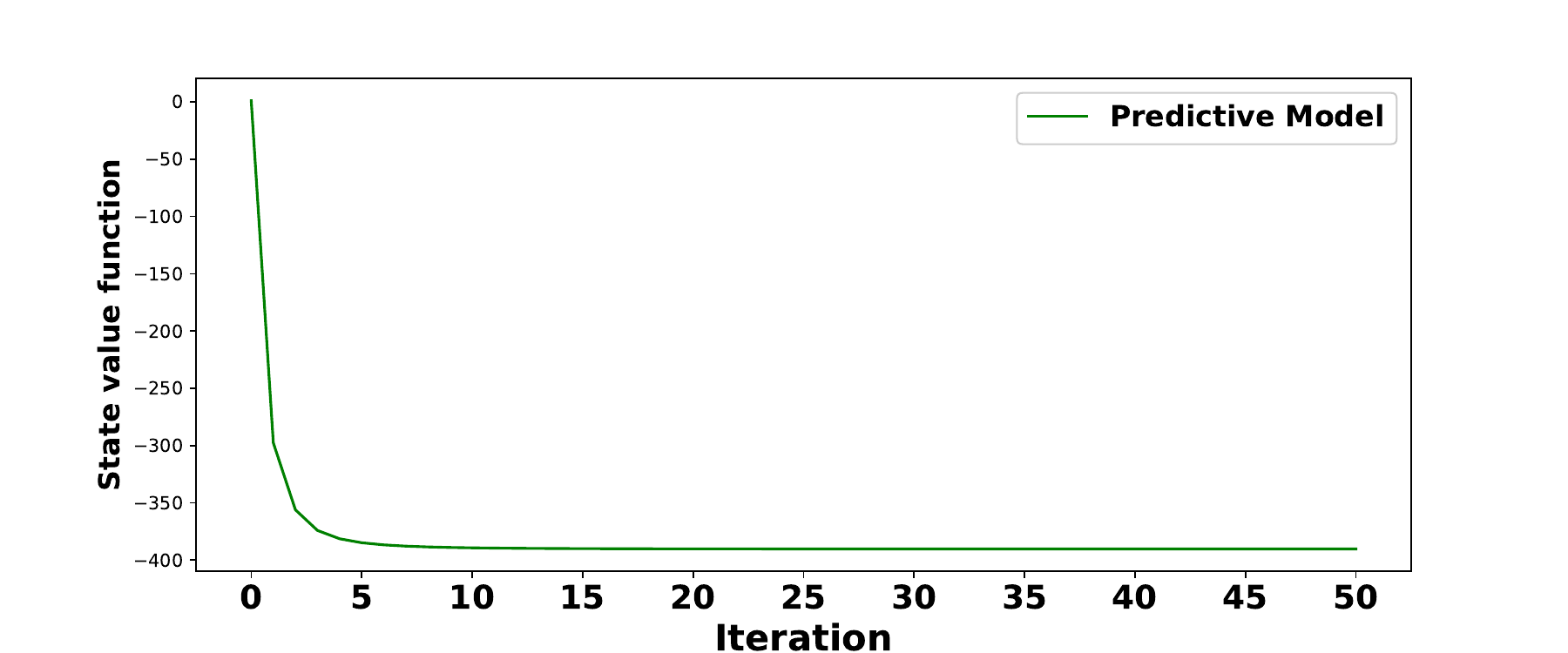}%
        \caption{Convergence of the state value \\ function over full state space.}%
        \label{fig:qlearn1}%
    \end{subfigure}\hfill%
    \begin{subfigure}{.50\columnwidth}
        \includegraphics[width=\columnwidth, height=5cm]{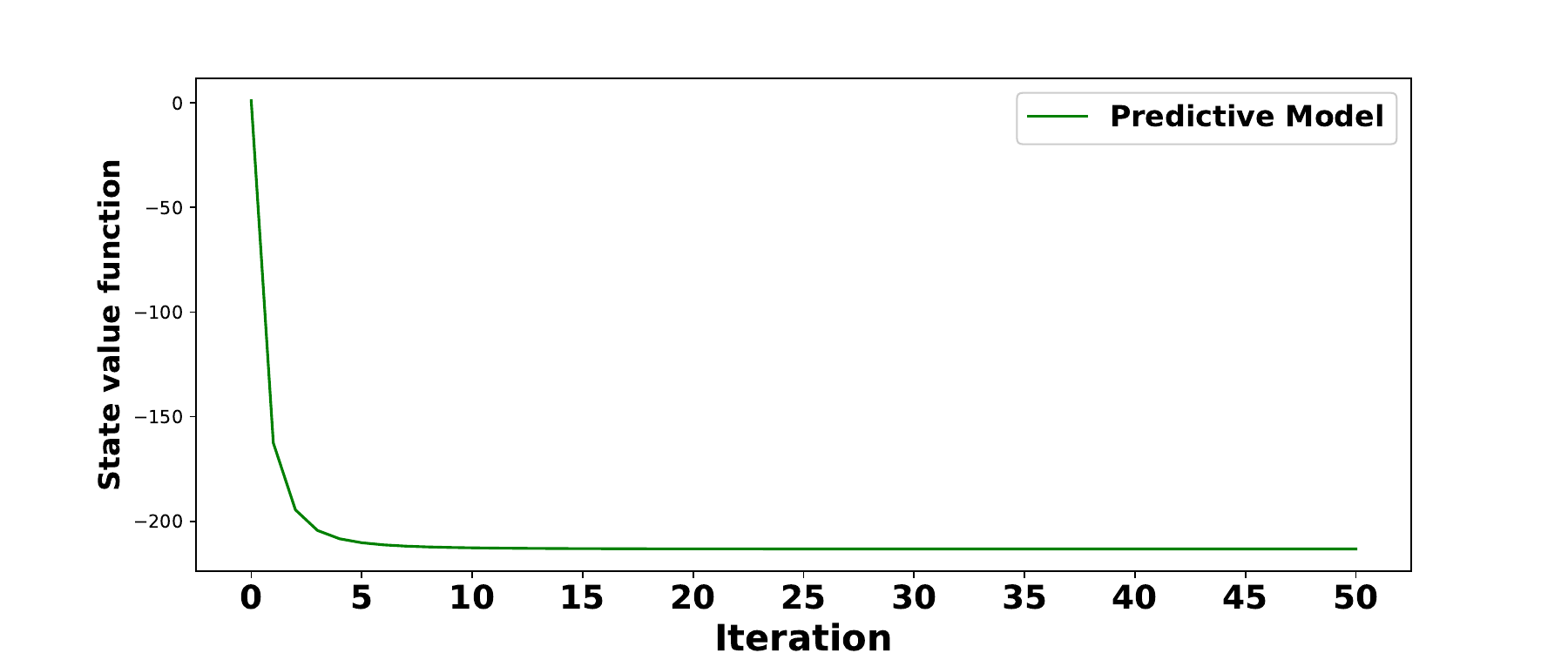}%
        \caption{Convergence of the state value function \\ over compact state space.}%
        \label{fig:qlearn2}%
    \end{subfigure}\hfill%
    \caption{Convergence of the state value function under various state spaces over network mobility.}%
    \label{fig:qlearn}%
\end{figure*}

\begin{figure*}[h!]%
    \centering
    \begin{subfigure}{.50\columnwidth}
        \includegraphics[width=\columnwidth, height=5cm]{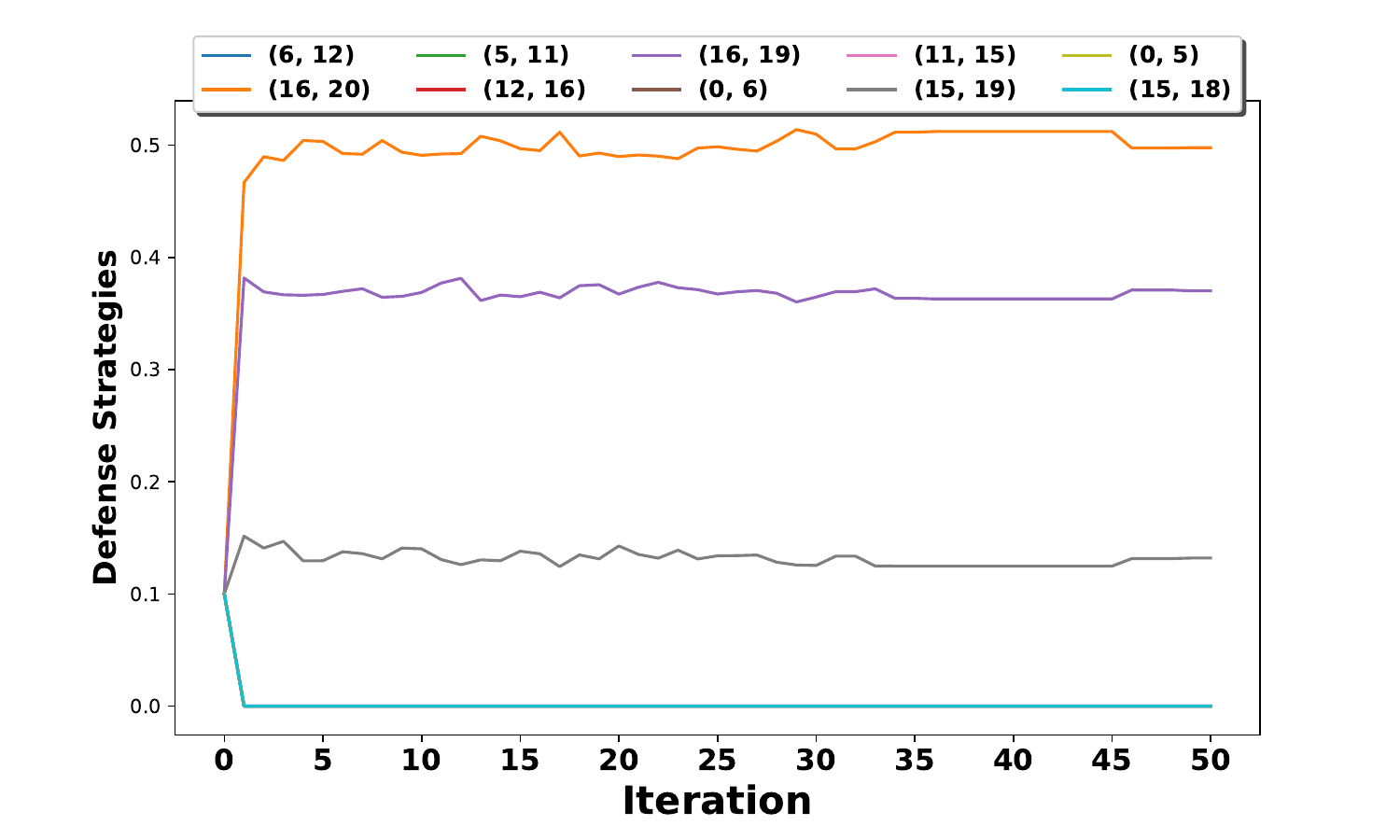}%
        \caption{Defender strategy convergences.}%
        \label{fig:defconv}%
    \end{subfigure}\hfill%
    \begin{subfigure}{.50\columnwidth}
        \includegraphics[width=\columnwidth, height=5cm]{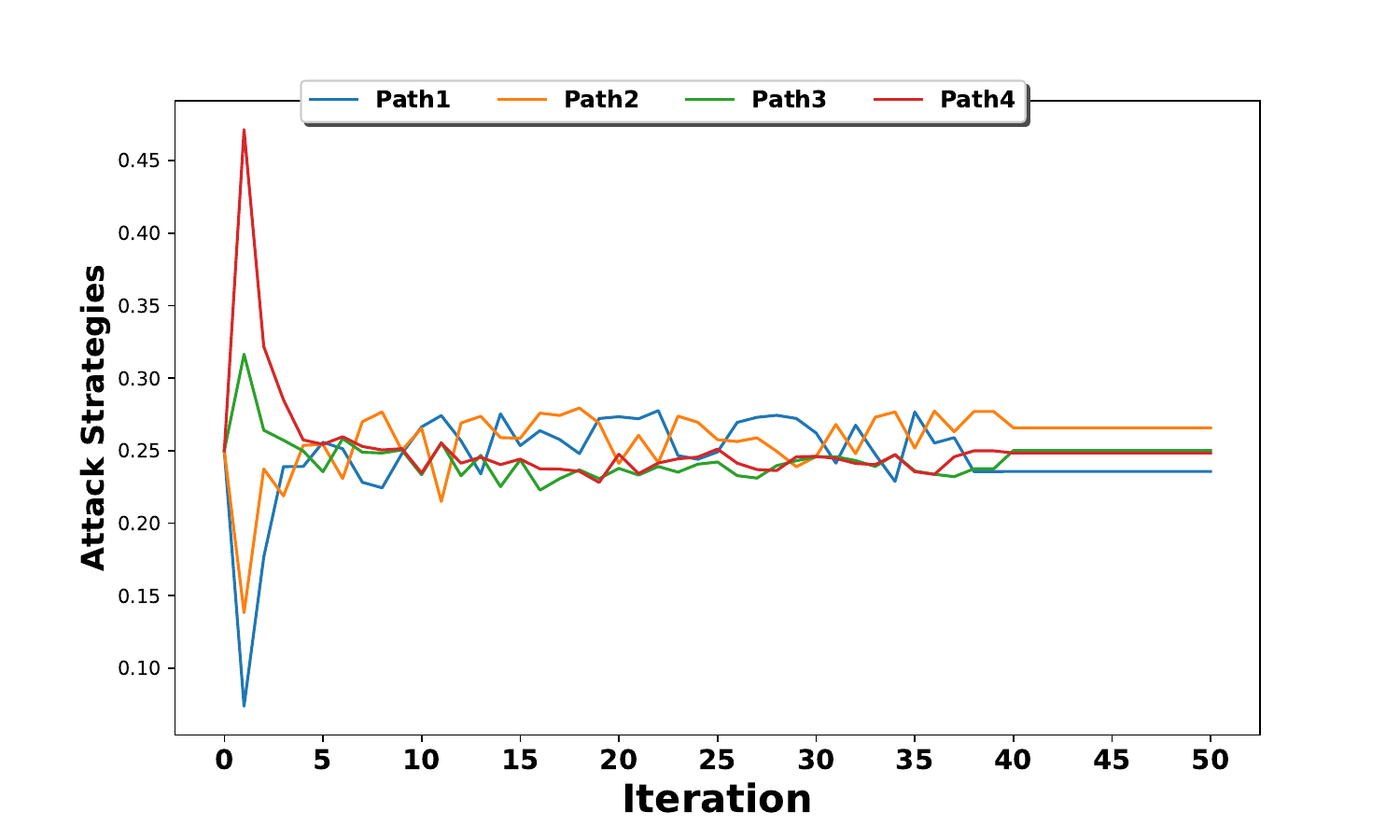}%
        \caption{Attacker strategy convergences.}%
        \label{fig:attconv}%
    \end{subfigure}\hfill%
    \caption{Defender and attacker strategy at a sample state converges to a mixed NE strategy at full state space.}%
    \label{fig:statconv}%
\end{figure*}

\textbf{Convergence:} In Fig.~\ref{fig:qlearn}, we illustrate the convergence of the value function for a sample state $S_0$ in both the full and compact state formulation. The value function reaches convergence in less than 15 iterations. In each iteration, $Q(s,a_d,a_a)$ is updated, and the value of the state updates based on that. The convergence value of the state $S_0$ is different for full and compact state space as they have different numbers of states, but both signify that incorporating mobility info in player strategies is always beneficial in tactical networks. This is associated with the convergence of the corresponding defense and attack strategies, as shown in Fig. ~\ref{fig:defconv} and ~\ref{fig:attconv}. 

Defender's strategy reaches convergence to the Nash Equilibrium (NE) policy depicted in Fig. ~\ref{fig:defconv}, where three actions have non-zero probability values while seven actions have zero probability values. Defender mixes between these 3 locations to deploy honeypot based on node significance and potential future rewards afterward. Simultaneously, the attacker's strategy converges to a NE policy as shown in Fig. ~\ref{fig:attconv}. Attacker selects (mixing randomly) among paths 2,3,4 with the presence of mobility as selecting between these paths maximizes the attacker expected reward as well as assists the attacker in successfully conducting lateral movement attacks and reaching the target, while playing path 1 with zero probability. 

\begin{figure*}[h!]%
    \centering
    \begin{subfigure}{.50\columnwidth}
        \includegraphics[width=\columnwidth, height=5cm]{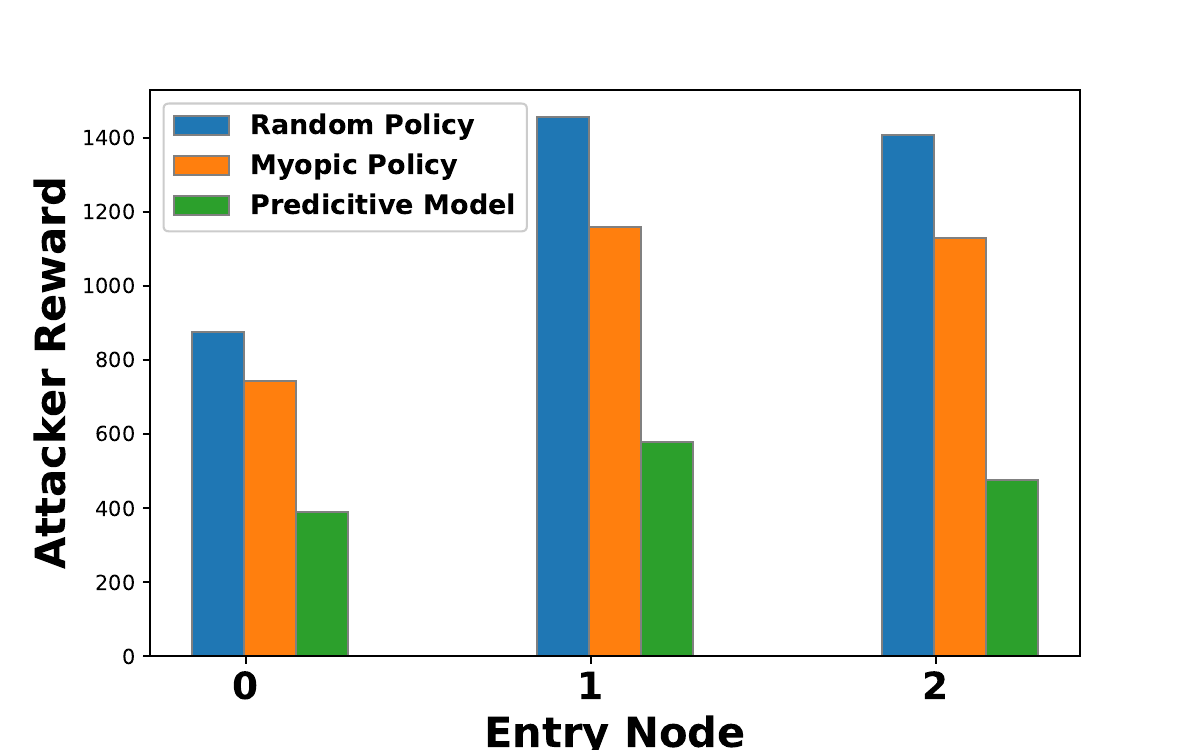}%
        \caption{Attacker reward over various\\ defender policy in full state space.}%
        \label{fig:p}%
    \end{subfigure}\hfill%
    \begin{subfigure}{.50\columnwidth}
        \includegraphics[width=\columnwidth, height=5cm]{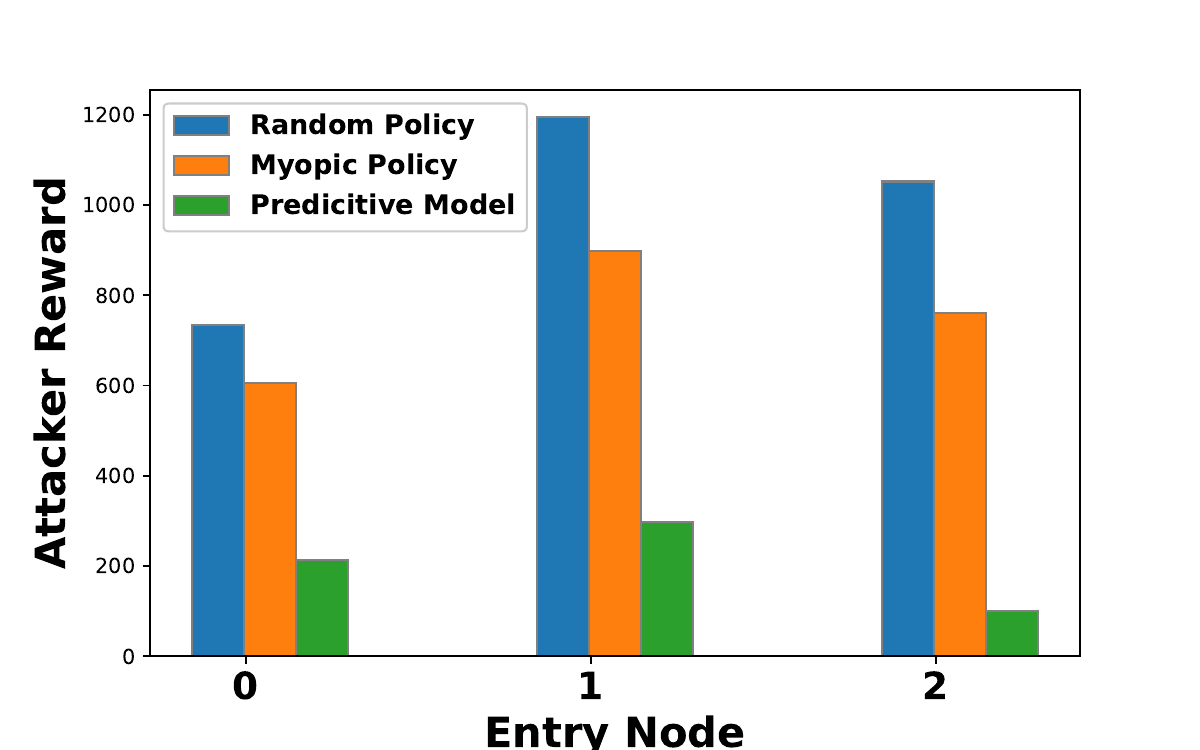}%
        \caption{Attacker reward over different\\ defender policy in compact state space.}%
        \label{fig:p1}%
    \end{subfigure}\hfill%
    \caption{Attacker reward for different defender policies over entry nodes under various state spaces.}%
    \label{fig:policy}%
\end{figure*}

\begin{figure*}[h!]%
    \centering
    \begin{subfigure}{.50\columnwidth}
        \includegraphics[width=\columnwidth, height=5cm]{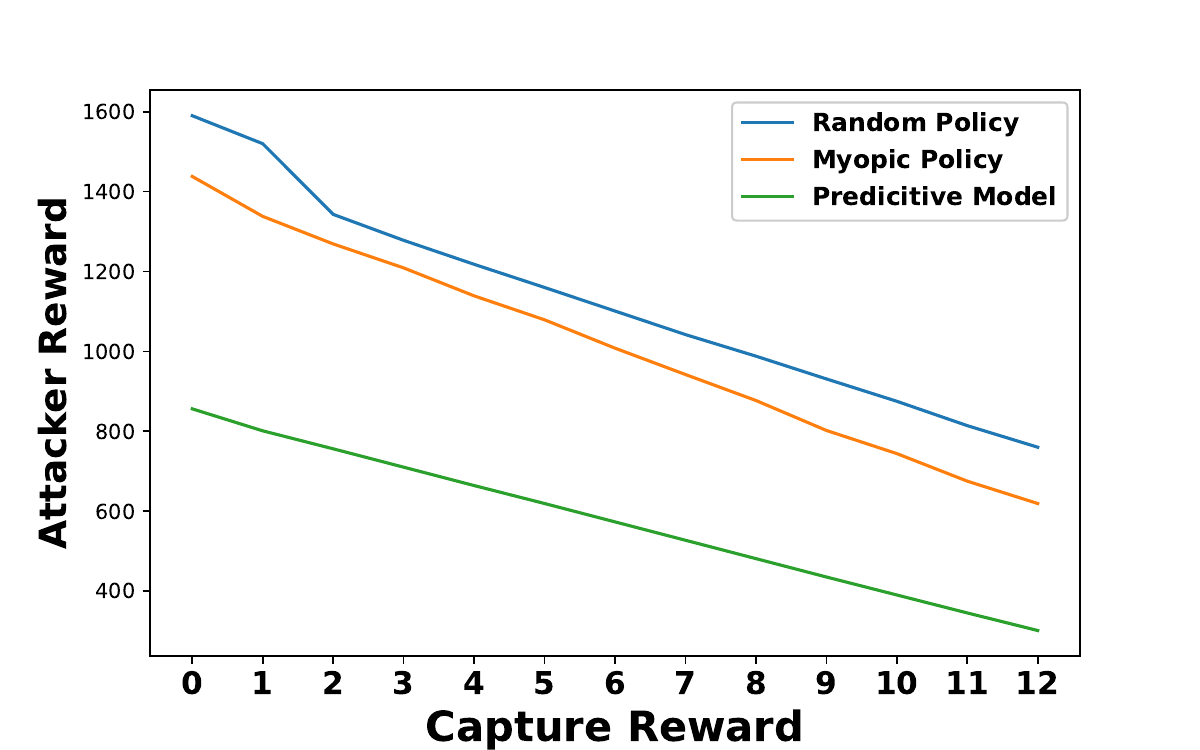}%
        \caption{Attacker reward for different defender\\ policy over capture values.}%
        \label{fig:cap}%
    \end{subfigure}\hfill%
    \begin{subfigure}{.50\columnwidth}
        \includegraphics[width=\columnwidth, height=5cm]{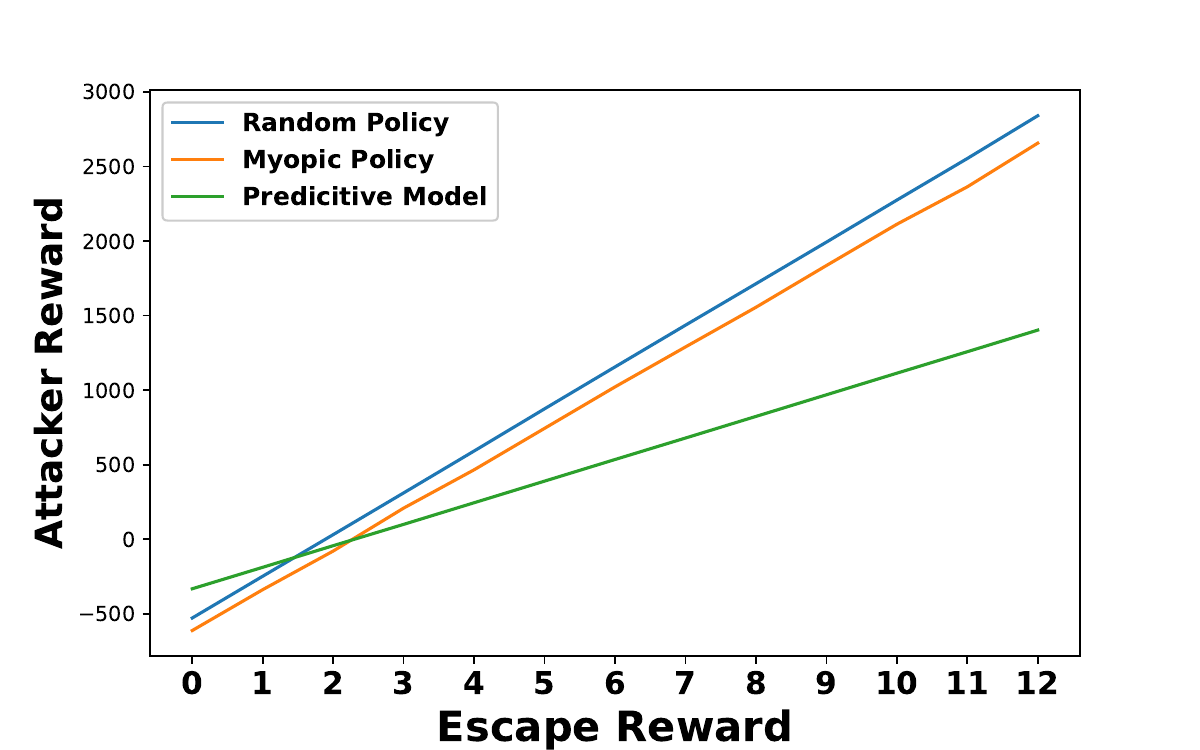}%
        \caption{Attacker reward for various defender\\ policy over escape values.}%
        \label{fig:esc}%
    \end{subfigure}\hfill%
    \caption{Attacker reward at a sample state for different capture and escape values.}%
    \label{fig:capesc}%
\end{figure*}

\textbf{Performance:} In Fig.~\ref{fig:policy}, we analyze the attacker's reward under different defender policies, considering random, myopic, and our predictive model policies, for both the full state and compact state formulations. Both figures (a, b) demonstrate that the predictive model policy outperforms both other policies, weak deception policies result in increasing the attacker reward. The variations in the attacker's reward across different entry nodes can be attributed to changes in the attacker and defender action spaces, the availability of new attack paths, as well as the intermediate node spaces showing that entry node 1 is more impactful when compromised.

 Fig. ~\ref{fig:cap} demonstrates when the capture reward increases, the attacker's overall reward decreases across various defender policies. It is worth noting here that the attacker has no option to completely back off assuming a persistent attacker. On the other hand, Fig. ~\ref{fig:esc} shows that as the attacker's escape reward increases, the attacker's reward continuously increases across different defender policies. It is important to note that the negative values of the escape rewards (0-2) do not contradict the results, as they signify a decrease in the attacker's reward. Overall, it shows again that the myopic policy outperforms the random policy while adhering to the predictive model policy results in the most significant reduction in the attacker's reward.

\begin{figure}[h!]
    \centering
    \includegraphics[width=10cm, height=5cm]{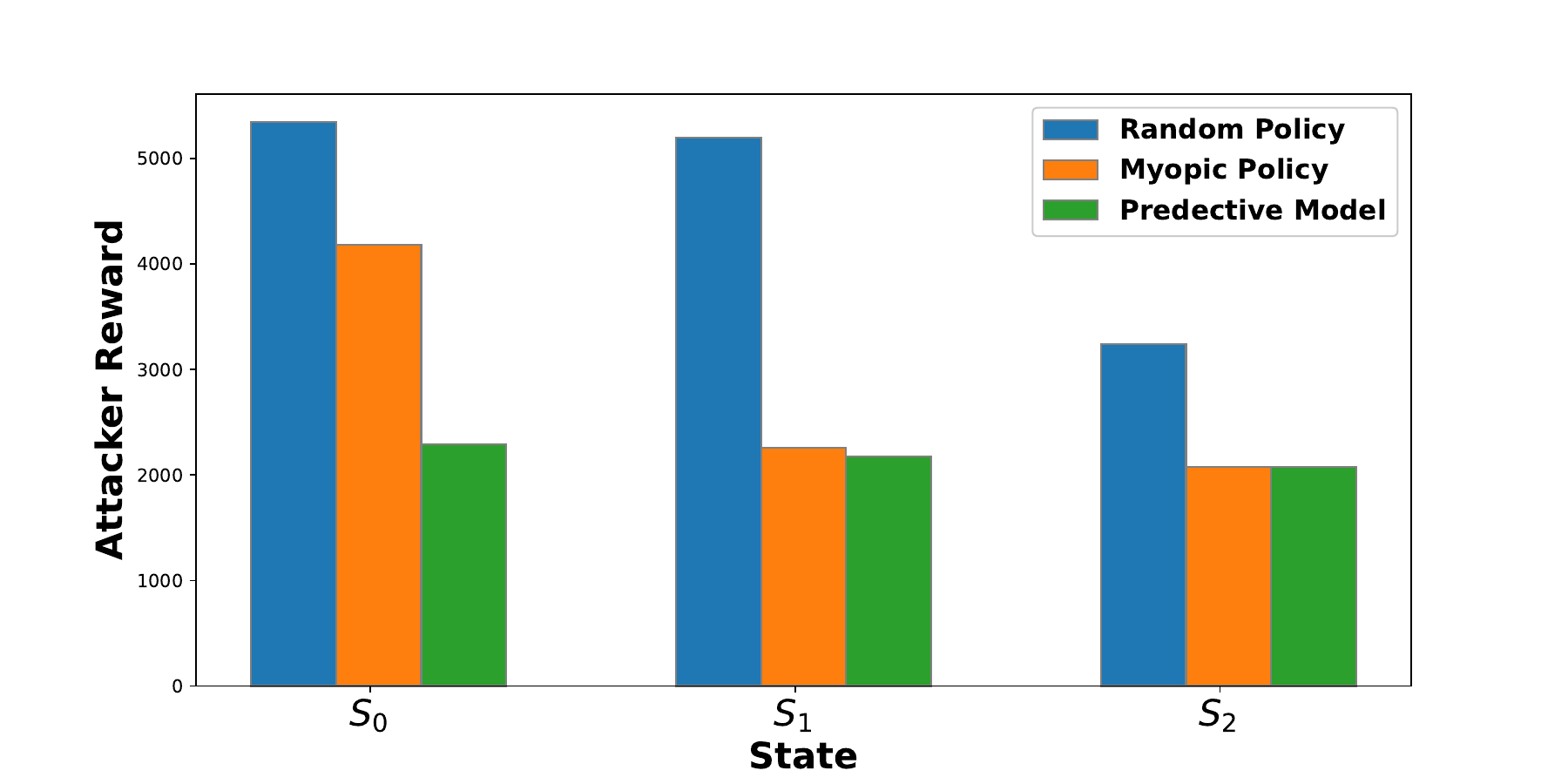}
    \caption{Attacker reward over various depth nodes in 50 nodes network.}
    \label{fig:policy50}
\end{figure}

Fig.~\ref{fig:policy50} illustrates the attacker reward for different points of time at different deception policies for a 50-node network. For the node at $S_2$ (terminal state), the attacker reward remains the same for both the myopic and predictive model policies, as there are no future states reward to be considered onward from that state. At $S_1$, the attacker reward is slightly higher for the defender's myopic policy compared to the predictive model policy. However, the improvement between the two policies becomes more significant at $S_0$, highlighting the importance of accounting for future evolution due to mobility instead of responding to sudden changes in network connectivity in a reactive fashion. Consequently, this demonstrates that proactive deception provides an advantage to the defender, highlighting the potential impact of implementing a moving target defense strategy alongside deception. Such an approach can disrupt the reconnaissance information that an adversary may gather regarding future transitions in the network.

\begin{figure*}[h!]%
    \centering
    \begin{subfigure}{.50\columnwidth}
        \includegraphics[width=\columnwidth, height=5cm]{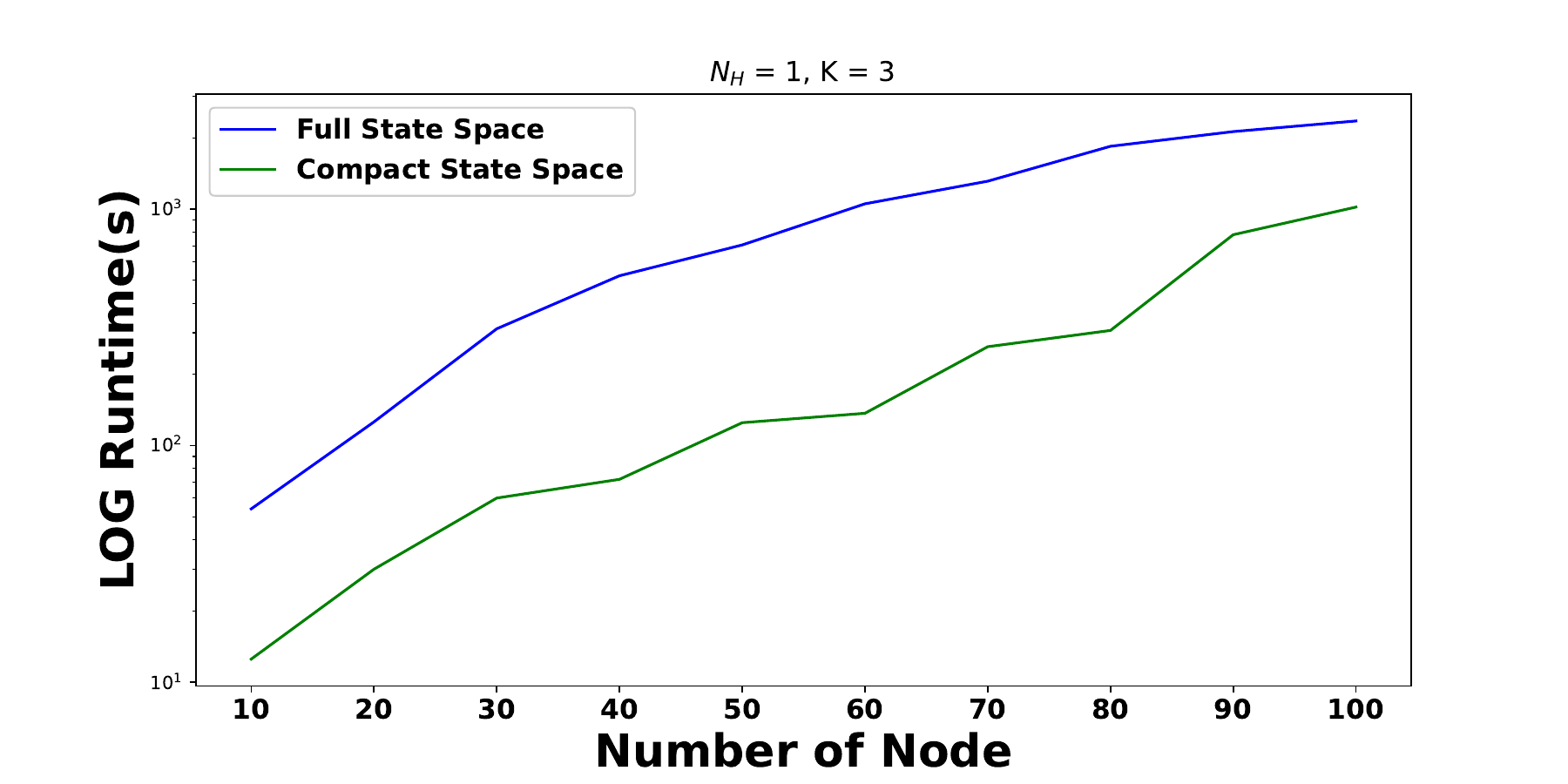}%
        \label{fig:sca1}%
    \end{subfigure}\hfill%
    \begin{subfigure}{.50\columnwidth}
        \includegraphics[width=\columnwidth, height=5cm]{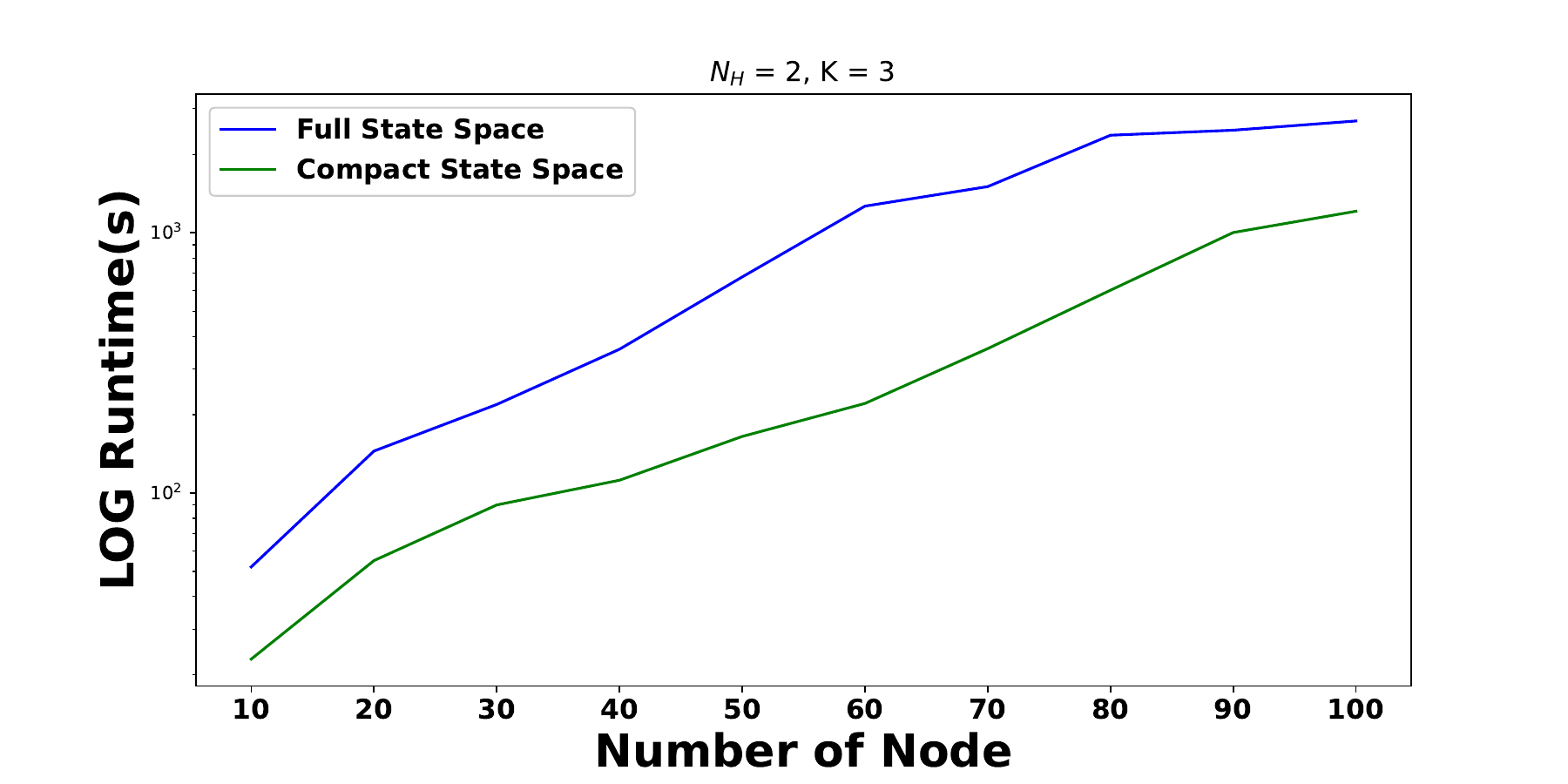}%
        \label{fig:sca2}%
    \end{subfigure}\hfill%
    \caption{Scalability in the number of nodes in the network under full and compact state spaces.}%
    \label{fig:sca}%
\end{figure*}

\textbf{Scalability:} We generate different Watts-Strogatz graphs by NetworkX library and demonstrate the overall complexity of different formulations under various parameters in Fig.~\ref{fig:sca}. In this context, the variable $H$ represents the number of honeypots, and $K$ represents the average number of nearest neighbors to each node. In Fig.~\ref{fig:sca} the running time increases exponentially for the full state-space consideration due to the involvement of the intermediate node space in mobility. 
In addition, increasing the honeypot budget increases, which enlarges the defender's action space, and also increases the runtime. However, considering the compact state-space formulation results in $44\%$ reduction in runtime compared to full state space for a network of 100 nodes, and $H=2$.

\section{Conclusion and Future Work}\label{sec:con}
In this paper, we present a dynamic game model involving a defender and an attacker to analyze the effectiveness of cyber deception in mitigating lateral move attacks in the presence of network mobility. Our proposed approach incorporates a multistage node removal technique and a predictive model to determine stationary mixed strategies by solving the dynamic game. We evaluate the effectiveness of the deception method under different network settings. Our numerical results and simulations it is evidence that cyber deception  strategies were able to mitigate the attack impact.
Finally, we showcase the scalability of our approach in terms of network size under mobility and provide a compact state space. 

There are several potential extensions to this work. One avenue for further exploration is optimizing the node removal process through parallel computation, which can significantly improve speed. However, addressing the inter-dependency between states in parallel calculations requires further investigation. Additionally, network mobility can be modeled as a moving target defense, involving the implementation of various strategies and techniques to make the network more dynamic and unpredictable for potential attackers. This can be achieved through careful planning and coordination. We also would like to consider the erroneous topology case, where the defender can obtain a noisy version of the topology. Quantifying the noise and leveraging stochastic attack graph models can help address this case which is part of our ongoing research. Another obvious extension is using function approximation such as deep Q-learning to scale to much larger/more complex games.

\bibliographystyle{unsrt}
\bibliography{reference}
\end{document}